\documentclass[journal=jacsat,manuscript=article]{achemso}
\setkeys{acs}{articletitle = true} 
\usepackage{xr} 
\usepackage{longtable}
\usepackage{multirow}
\usepackage{amsmath}
\usepackage{amssymb}
\usepackage{subfigure}
\usepackage[usenames,dvipsnames]{color}
\SectionNumbersOn
\author{Oliver T. Unke}
\affiliation{Department of Chemistry, University of Basel, Klingelbergstrasse 80
, CH-4056 Basel, Switzerland.}
\author{Markus Meuwly}
\affiliation{Department of Chemistry, University of Basel, Klingelbergstrasse 80
, CH-4056 Basel, Switzerland.}
\email{m.meuwly@unibas.ch}
\date{\today}

\title{Machine Learning Potential Energy Surfaces}

\begin{document}
Machine Learning techniques can be used to represent high-dimensional
potential energy surfaces for reactive chemical systems. Two such
methods are based on a reproducing kernel Hilbert space representation
or on deep neural networks. They can achieve a sub-1 kcal/mol accuracy
with respect to reference data and can be used in studies of chemical
dynamics. Their construction and a few typical examples are briefly
summarized in the present contribution.
\begin{abstract}

\end{abstract}

\section{Introduction}
The potential energy surface (PES) - the way how atoms and molecules
interact with one another - contains all information necessary to
describe the structure and to follow the dynamics of molecular
systems. Characterizing, interpreting and understanding the dynamics
of homogeneous and heterogeneous systems at a molecular level is a
formidable task.\cite{MM.rev.sd:2017} Typical processes of interest
are chemical reactions or functional motions in proteins. For chemical
reactions the relevant process, i.e. bond breaking or bond formation,
occurs on the femtosecond time scale whereas typical rates for
solution phase reactions are in the range of 1 s$^{-1}$. In other
words, during $\sim 10^{15}$ vibrational periods energy is
redistributed in the system until sufficient energy has accumulated
along the preferred ``progression coordinate'' for the reaction to
occur. Another example, freezing and phase transitions in water are
entirely governed by intermolecular interactions. Describing them at
sufficient detail has been found extremely challenging and a complete
understanding of the phase diagram or the structural dynamics of
liquid water is still not available.\cite{water.rev:2016}\\

\noindent
Computing an experimentally observable quantity, such as a rate,
infrared spectrum, or a diffusion coefficient, requires means to
determine the total energy of the system accurately and
efficiently. For MD simulations not only energies but also forces are
required to propagate the equations of motion. One way is to solve the
electronic Schr\"odinger equation for every configuration $\vec{x}$ of
the system. However, there are limitations which are due to the
computational approach {\it per se}, e.g. the speed and efficiency of
the method or due to practical aspects of quantum chemistry such as
accounting for the basis set superposition error, the convergence of
the Hartree-Fock wavefunction to the desired electronic state for
arbitrary geometries, or the choice of a suitable active space
irrespective of molecular geometry for problems with multi-reference
character, to name a few. Improvements and future avenues for making
QM-based approaches even more broadly applicable have been recently
discussed.\cite{qiang.jcp:2016} For problems that require extensive
conformational sampling or sufficient statistics purely QM-based
dynamics approaches are still impractical.\\

\noindent
A promising use of QM-based methods are mixed quantum
mechanics/molecular mechanics (QM/MM) treatments which are
particularly popular for biophysical and biochemical
applications.\cite{senn:2009} Here, the system is decomposed into a
``reactive region'' which is treated with a quantum chemical (or
semiempirical) method and an environment described by an empirical
force field. Such a decomposition considerably speeds up simulations
such that even free energy simulations in multiple dimensions can be
computed.\cite{cui:2016} One of the current open questions in such
QM/MM simulations is that of the size of the QM region required for
converged results which was recently considered for Catechol
O-Methyltransferase.\cite{kulik:2016}\\

\noindent
The most rigorous approach to reactive chemical dynamics is to use
fully dimensional, reactive PESs, which are usually only available for
low-dimensional systems. Specifically in the field of small molecule
reactions involving processes such as A+BC$\rightarrow$AB+C or
AB+CD$\rightarrow$A+BCD (or permutations thereof) involving a total
number of 3 to 8 atoms the construction of globally valid potential
energy surfaces is essential to make direct contact between
computations and experiments. The limitation on up to 8 atoms is
primarily owed to the fact that accurately solving the nuclear
Schr\"odinger equation for systems involving a larger number of nuclei
remains an unsolved problem although considerable progress has been
made over the past several years.\cite{bowman:2010,zhang:2017}\\

\noindent
One of the earliest efforts to determine outcomes of chemical
reactions from dynamics studies is the H+H$_2$ atom exchange
reaction. Using classical molecular dynamics (MD) and a modified
London-Eyring-Polanyi-Sato (LEPS)
surface,\cite{karplus.h3:1964,leps1,leps2} the differential cross
sections for the D+H$_2$ reaction were calculated.\cite{karplus:1964}
Some 10 years later the results of these quasi-classical simulations
were almost quantitatively confirmed at room temperature by a full
quantum treatment.\cite{schatz:1976}\\

\noindent
High-level {\it ab initio} calculations have become more accurate and
more
efficient.\cite{white1996linear,adler2007simple,knizia2009simplified}
Along with the ever-increasing computational power it is now possible
to compute thousands of energies at various geometries for small
polyatomic systems within chemical accuracy (0.5
kcal/mol).\cite{bokhan2008implementation} With the availability of
routine electronic structure calculations using large basis sets and
accurate ways for approximately solving the electronic Schr\"odinger
equation the problem shifted to {\it representing} the computed,
discrete points on the PES. This is in particular needed when running
MD simulations (classical or quasiclassical trajectory (QCT)
simulations) where the total energy and the derivatives at arbitrary
points are required. Similarly, quantum wavepacket calculations
require analytical forms of the PES whereas for collocation methods it
is sometimes sufficient to work with a discrete, precomputed set of
energies. Nevertheless, it is nowadays standard to represent
nonreactive and reactive PESs in a manner that allows to evaluate it
at arbitrary points, given pointwise information from {\it ab initio}
calculations only. In addition, the final representation should be
computationally efficient in order to run statistically significant
numbers of (reactive) trajectories in QCT
simulations.\cite{yosa:2011,mm12n4p,castro2014computational}\\

\noindent
A direct way for obtaining an analytical representation of a PES is to
use a parametrized functional form\cite{hutson:1990,aguado1992new} and
fit the parameters to a set of \textit{ab initio} data using linear or
non-linear least squares procedures.\cite{law1997nolls} For the
specific case of van der Waals molecules such an approach has been
very successful, in particular for applications in
spectroscopy.\cite{hutson:1990,hutson.arco2:1996} Although such
approaches have demonstrated to achieve root mean squared errors
(RMSEs) within chemical accuracy,\cite{boothroyd1996refined} choosing
a functional form requires human intuition and the fitting itself can
be tedious and time-intensive.\cite{MM.heh2:1999,water.avoird:2000}\\

\noindent
Over the past years alternative interpolation techniques including the
modified Shepard
interpolation\cite{franke1980smooth,nguyen1995dual,bettens1999learning},
the moving least squares
method\cite{lancaster1981surfaces,ischtwan1994molecular,dawes2008interpolating},
permutation invariant
polynomials\cite{cassam2008symmetry,braams2009permutationally,paukku2013global}
or neural network
approaches\cite{sumpter1992potential,bowman2010ab,jiang2016potential}
have been used to obtain multi-dimensional reactive
PESs.\cite{jordan1995utility,jordan1995convergence,skokov1998accurate,collins2002molecular,duchovic2002potlib,zhang2006ab,li2012communication}
The total number of coefficients $D_{a_1 \cdots a_m}$ grows rapidly
with the number of atoms. For total polynomial order of 5 and 4 atoms
(molecule ABCD with 4 different atoms) there are 462 such coefficients
and for 5 atoms (molecule ABCDE) there are 3003 of them. Thus, the
number of reference energies that need to be determined depends on the
number of distinct coefficients for the fitting problem to be
well-defined. Common to all these approaches is the fact that they
primarily minimize the RMSE with respect to the training data. In this
respect they are similar to parametrized fits..\\

\section{Reproducing Kernel Hilbert Space (RKHS) PES}
Machine-learning (ML) methods allow to estimate an unknown function
value using a model that was ``trained" with a set of known
data.\cite{rupp2015machine} For intermolecular interactions, Rabitz
and
coworkers\cite{ho1996general,hollebeek1997fast,hollebeek1999constructing}
have popularized the use of reproducing kernel Hilbert space (RKHS)
theory\cite{aronszajn1950theory} that allows to construct a PES from a
training set based on \textit{ab initio} reference data. Such an
approach is typically referred to as kernel ridge regression (KRR) in
the ML community.\cite{hofmann2008kernel,rupp2015machine} The RKHS
method has been successfully applied e.g. for constructing PESs for
NO+O\cite{castro2014computational}, N$_2^+$ +
Ar\cite{unke2016collision} or H$_2$O.\cite{ho1996global} A combination
of expanding the PES in spherical harmonics for the angular
coordinates and reproducing kernels for the radial coordinates has
been explored for H$_2^+$--He\cite{MM.heh2:1999} and is now also used
for larger systems.\cite{avoird:2005,avoird:2006}\\

\noindent
To further automatize this process, dedicated computer code has been
made available that generates the interpolation (and meaningful
extrapolation) of the PES along with all required parameters
automatically from gridded {\it ab initio} data.\cite{MM.rkhs:2017}
The theory of reproducing kernel Hilbert spaces asserts that for given
values $f_i = f(\mathbf{x_i})$ of a function $f(\mathbf{x})$ for $N$
training points $\mathbf{x}_i$, $f(\mathbf{x})$ can always be
approximated as a linear combination of kernel products
\cite{scholkopf2001generalized}
\begin{equation}
\widetilde{f}(\mathbf{x}) = \sum_{i = 1}^{N} c_i K(\mathbf{x},\mathbf{x}_i)
\label{eq:RKHS_function}
\end{equation}
Here, the $c_i$ are coefficients and $K(\mathbf{x},\mathbf{x'})$ is
the reproducing kernel of the RKHS. The coefficients $c_i$ satisfy the
linear relation
 \begin{equation}
 f_j = \sum_{i = 1}^{N} c_i K_{ij}
 \label{eq:RKHS_coefficient_relation}
 \end{equation}
with the symmetric, positive-definite kernel matrix $K_{ij} =
K(\mathbf{x}_i,\mathbf{x}_j)$ and can therefore be calculated from the
known values $f_i$ in the training set by solving
Eq. \ref{eq:RKHS_coefficient_relation} for the unknowns $c_i$ using,
e.g. Cholesky decomposition.\cite{golub2012matrix} With the
coefficients $c_i$ determined, the function value at an arbitrary
position $\mathbf{x}$ can be calculated using
Eq.~\ref{eq:RKHS_function}. Derivatives of $\widetilde{f}(\mathbf{x})$
of any order can be calculated analytically by replacing the kernel
function $K(\mathbf{x},\mathbf{x'})$ in Eq. \ref{eq:RKHS_function}
with its corresponding derivative.\\

\noindent
For practical applications in chemical physics it is also of interest
to mention that it is possible to handle incomplete grids with an RKHS
interpolation.\cite{hollebeek1999constructing} This is important
because even for triatomic systems it is possible that converging
quantum chemical calculations turns out to be difficult for certain
geometries. Under these circumstances the grid of target energies
contains a ``hole''.\\

\noindent
The explicit form of the multi-dimensional kernel function
$K(\mathbf{x},\mathbf{x'})$ is chosen depending on the problem to be
solved. In general, it is possible to construct $D$-dimensional
kernels as tensor products of one-dimensional kernels $k(x,x')$
\begin{equation}
K(\mathbf{x},\mathbf{x'}) = \prod_{d=1}^{D} k^{(d)}(x^{(d)},x'^{(d)})
\label{eq:multidimensional_kernel}
\end{equation}
For the kernel functions $k(x,x')$ it is possible to encode physical
knowledge, in particular about its long range behaviour. Explicit
radial kernels include the reciprocal power decay kernel
\begin{equation}
k_{n,m}(x,x') = n^2 x_{>}^{-(m+1)}\mathrm{B}(m+1,n)_2\mathrm{F}_1\left(-n+1,m+1;n+m+1;\dfrac{x_{<}}{x_{>}}\right)
\label{eq:reciprocal_power_kernel}
\end{equation}
 or the exponential decay kernel
\begin{equation}
k_{n}(x,x') = \dfrac{n\cdot n!}{\beta^{2n-1}} \mathrm{e}^{-\beta x_{>}}\sum_{k=0}^{n-1}\frac{(2n-2-k)!}{(n-1-k)!k!}\left[\beta(x_{>}-x_{<})\right]^k
\label{eq:exponential_kernel}
\end{equation}
where $x_{>}$ and $x_{<}$ are the larger and smaller of $x$ and $x'$
and the integer $n$ determines the smoothness. In
Eq. \ref{eq:reciprocal_power_kernel} the parameter $m$ is the
long-range decay of the dominant intermolecular interaction
(e.g. $m=6$ for dispersion), $\mathrm{B}(a,b)$ is the beta function
and $_2\mathrm{F}_1(a,b;c;d)$ is the Gauss hypergeometric
function.\\

\section{Deep Neural Networks}
Artificial neural networks
(NNs)\cite{mcculloch1943logical,kohonen1988introduction,abdi1994neural,bishop1995neural,clark1999neural,ripley2007pattern,haykin2009neural}
are a popular class of ML algorithms to tackle computationally
demanding
problems.\cite{hinton2012deep,lawrence1997face}. Specifically, NNs
have been used previously to fit PESs for molecular systems in the
spirit of many-body
expansions\cite{manzhos2006random,manzhos2007using,malshe2009development,handley2010potential}. While
being accurate, they typically involve a large number of individual
NNs (one for each term in the many-body expansion), making the method
scale poorly for large systems. An alternative approach, known as
high-dimensional NN (HDNN) first applied to bulk
silicon,\cite{behler2007generalized,behler2011neural} decomposes the
total energy of a system into atomic contributions, which is
appealing, because ``energy'' is an extensive property and it allows
to use the same network to systems of different size.\\

\noindent 
Deep tensor NN (DTNN)\cite{schutt2017quantum} are a conceptually
different approach and allow to reuse the same NN to predict energies
of systems with different composition across chemical space. Similar
to HDNNs, DTNN accumulates atomic energy contributions to predict the
total energy $E_{\rm tot}$.\\

\noindent
In an artificial neural network
(ANN)\cite{mcculloch1943logical,kohonen1988introduction,abdi1994neural,bishop1995neural,clark1999neural,ripley2007pattern,haykin2009neural}
neurons transform an input vector $\mathbf{x}$ of dimension $n$ to an
output vector $\mathbf{y}$ of dimension $m$ through a transformation
\begin{equation}
\mathbf{y} = \mathbf{W}\mathbf{x} + \mathbf{b}
\label{eq:dense_layer}
\end{equation}
Here, $\mathbf{W}$ is an $n \times m$ matrix containing weights and
the biases $\mathbf{b}$ are an $n-$vector whose both entries are
parameters.  A single dense layer NN can only represent linear
relations between input and output. In order to model a non-linear
relationship at least two dense layers are required and combined with
a non-linear activation function $\sigma$ (e.g. a sigmoid function)
\begin{eqnarray}
\label{eq:single_layer_hidden}
\mathbf{h} =& \sigma\left(\mathbf{W_{1}}\mathbf{x} + \mathbf{b_{1}}\right)\\
\label{eq:single_layer_output}
\mathbf{y} =& \mathbf{W_{2}}\mathbf{h} + \mathbf{b_{2}}
\end{eqnarray}
Equations~\ref{eq:single_layer_hidden} and
\ref{eq:single_layer_output} are general function approximators with
which any input $\mathbf{x}$ can be mapped onto output $\mathbf{y}$ to
arbitrary precision, provided that the dimensionality of the ``hidden
layer'' $\mathbf{h}$ is large
enough.\cite{gybenko1989approximation,hornik1991approximation} As
such, NNs are a natural choice for representing a PES, i.e.\ a mapping
from chemical structure to energy. For an intermolecular PES the
output $\mathbf{y}$ is usually one-dimensional and represents the
energy.\\

\noindent
While \textit{shallow} NNs with a single hidden layer are in principle
sufficient to solve any learning task, in practice, \textit{deep} NNs
with multiple hidden layers are exponentially more
parameter-efficient.\cite{eldan2016power} In a deep NN, $l$ hidden
layers are stacked on top of each other, which map the input
$\mathbf{x}$ to an increasingly complex feature space. In the final
layer, the features $\mathbf{h_{l}}$ are combined to the output
$\mathbf{y}$. The parameters of the NN are initialized randomly and
then optimized by minimization of a loss function which quantifies the
difference between the output of the NN and the training data.\\

\noindent
{\it NN based on local descriptor:} Using a strictly local chemical
descriptor, a NN-based method tailored for accurate energy
evaluations, which can be applied to construct PESs for nonreactive
and reactive dynamics of chemically heterogeneous systems in the
condensed phase, has been introduced.\cite{MM.nn:2018} For this, a
descriptor has been built from representing the neighborhood density
of an atom as a linear combination of a product of a radial Gaussian
function and spherical harmonics. Training with 50000 structures from
the QM9 data set yields an RMSE of $0.98 \pm 0.04$ kcal/mol compared
with $1.37 \pm 0.01$ from DTNN.\cite{schutt2017quantum} The mean
average errors are $0.46 \pm 0.01$ kcal/mol compared with $0.92 \pm
0.01$ kcal/mol from DTNN and 0.59 from SchNet.\cite{schutt2017schnet}
Contrary to SchNet, this NN is more efficient because a local
descriptor is used and the network architecture is much simpler. As
forces can also be evaluated it is possible to train NN for reactive
MD simulations. However, it should be noted that developing NN-based
energy functions depends on the availability of extensive training
data.\\

\noindent
{\it NN based on learnable descriptor:} PhysNet, on the other hand,
learns the atom descriptors during training. It combines reusable
building blocks in a modular fashion to construct a DNN for predicting
atomic contributions to properties (such as energy) of a chemical
system composed of $N$ atoms based on atomic features $\mathbf{x}_{i}
\in \mathbb{R}^n$ (here, $n$ denotes the dimensionality of the feature
space). The features simultaneously encode information about nuclear
charge $Z$ and local atomic environment of each atom $i$ and are
constructed by iteratively refining an initial representation
depending solely on $Z_i$ through coupling with the feature vectors
$\mathbf{x}_{j}$ of all atoms $j\neq i$ within a cut-off radius
$r_{\rm cut}$. The {\it embedding layer} maps from a discrete object -
atomic numbers - to a vector of real numbers. The {\it residual block}
contains shortcut connection which help to maintain performance in
learning as training proceeds. The {\it interaction block} uses atom
positions in the features and is formulated in terms of pairwise
distances which ensures translational and rotational invariance, and
summation ensures permutational invariance. In this block, the
interaction between a central atom $i$ and its environment (atoms $j$
within a cutoff distance) is iteratively refined based on attention
masks that are biased towards $\exp{(-r_{ij})}$ to encode locality in
bonded chemical interactions. Finally, the {\it output block} computes
the result from each module by summation.\\

\section{Applications}
Dynamics studies of chemical reactions date back more than 50
years.\cite{karplus:1964} Since then the sophistication of the PESs
and the accuracy with which the nuclear dynamics can be followed have
continuously increased.\cite{bowman:2011} In the following a few
examples for using RKHS- and NN-based reactive PESs are briefly
discussed.\\

\subsection{Small Molecule Reactions}
\begin{figure}
\includegraphics[scale=0.67]{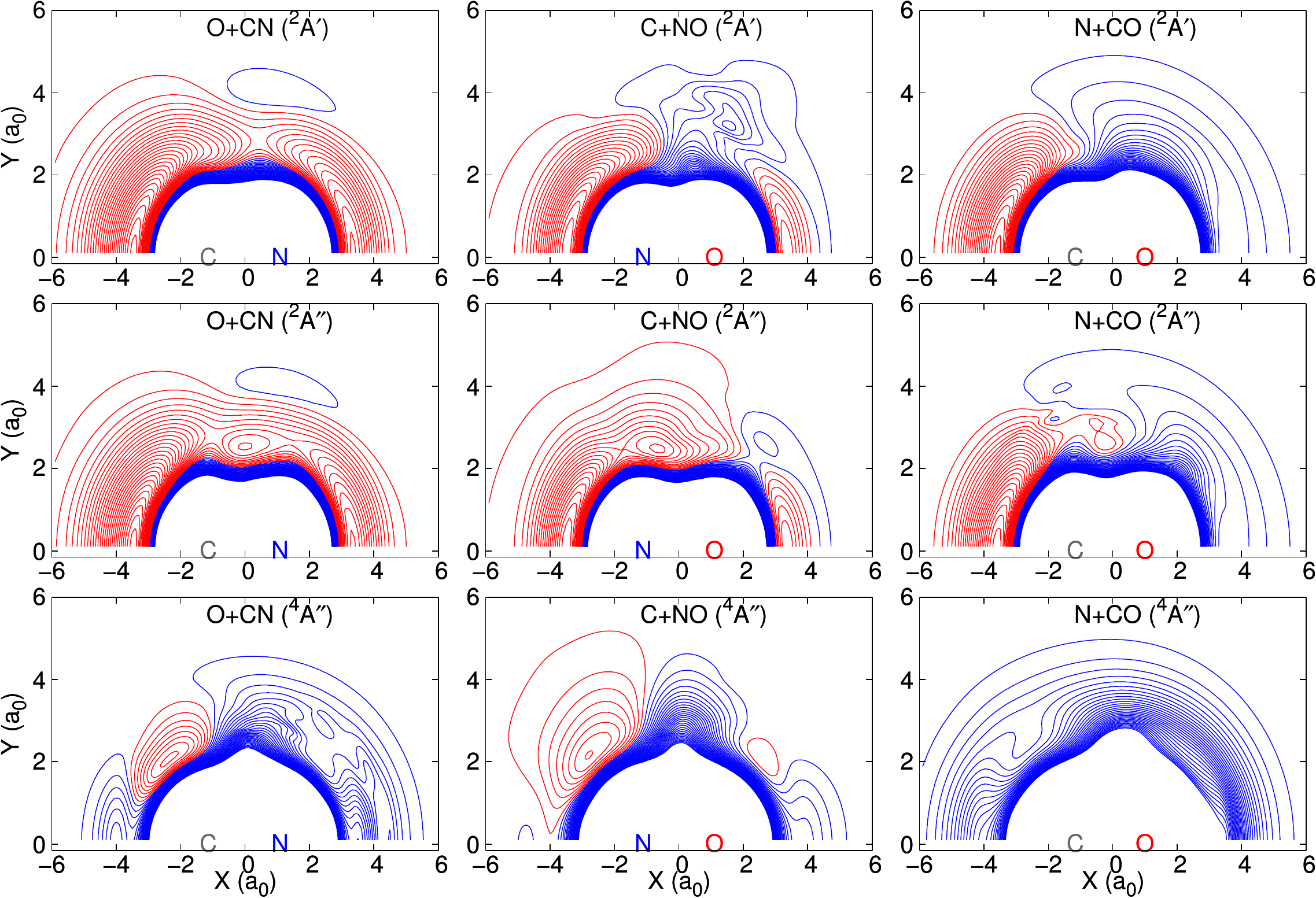}
\caption{Contour plots of the analytical PESs with the diatoms having
  bond distances fixed at their equilibrium distances. Spacing between
  the contour lines is 0.2 eV. The red lines correspond to negative
  energies (-0.1, -0.3, -0.5, .... eV) and the blue lines correspond
  to positive energies (0.1, 0.3, 0.5, .... eV). Zero of energies set
  to the energy of the atoms and equilibrium configurations of the
  diatoms ($i.\ e.,$ 2.234, 2.192 and 2.150 a$_0$ respectively for CN,
  NO and CO).}
\label{fig:figcno}
\end{figure}

Particularly interesting applications of reactive MD simulation
concern physical conditions that are difficult to study in laboratory
experiments, such as extremely high ($T \geq 10000$ K) temperatures as
they occur in explosions or in hypersonics.\cite{Millikan:63} As an
example, the C($^3$P) + NO(X$^2\Pi$) $\rightarrow$ O($^3$P) +
CN(X$^2\Sigma^+$), N($^2$D)/N($^4$S) + CO(X$^1\Sigma^+$) reactions of
the [CNO] system are briefly considered.\cite{MM.cno:2018} Figure
\ref{fig:figcno} shows the RKHS representations of the $^2$A$'$,
$^2$A$''$ and $^4$A$''$ electronic states based on more than 50000
    {\it ab initio} energies are calculated at the MRCI+Q/aug-cc-pVTZ
    level of theory. As can be appreciated some of the topographies of
    the PESs would be very demanding to be represented by a particular
    choice of a parametrized function, such as the $^2$A$''$ channel
    for N+CO. Under such circumstances an RKHS, data-driven ansatz is
    preferable.\\

\noindent
These PESs can subsequently be used in quasi-classical or quantum
studies. As an example, the rate for CN and CO formation from C($^3$P)
+ NO(X$^2\Pi$) $\rightarrow$ O($^3$P) + CN(X$^2\Sigma^+$) and
N($^2$D)/N($^4$S) + CO(X$^1\Sigma^+$) is compared. Without inclusion
of nonadiabatic effects the two experimentally determined rates are
underestimated by $\sim 25$ \% whereas including them leads to
quantitative agreement within error bars. As another finding it was
shown that the product vibrational state distribution of CN and CO
following the C+NO reaction from quasi-classical and quantum
simulations are very similar for different collisional energies.\\

\subsection{Reactive Dynamics using Neural Networks}
\begin{figure}
\includegraphics[width=0.75\textwidth]{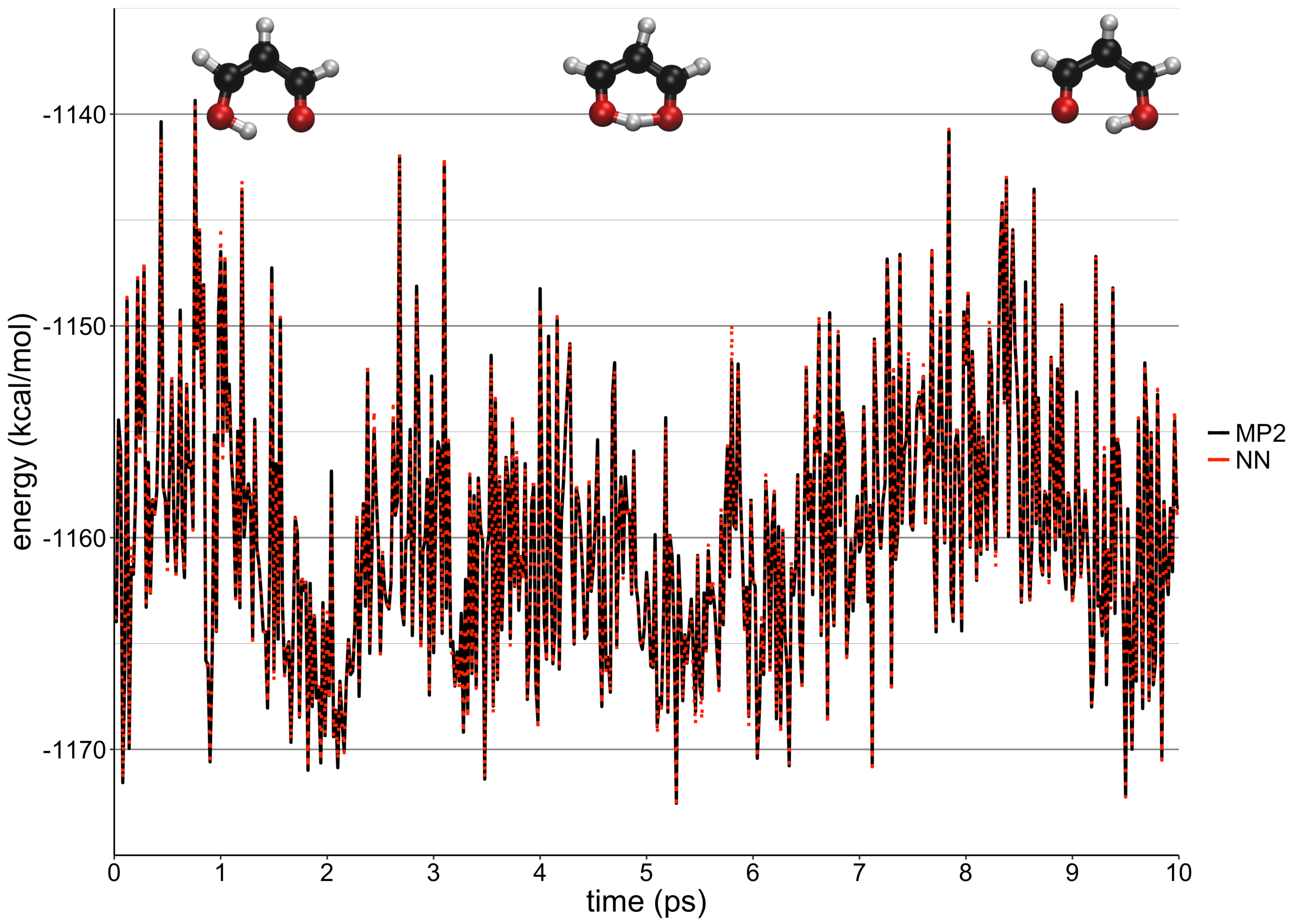}
\caption{First 10 ps of a MD trajectory of malonaldehyde with
  intramolecular H-transfer. \textit{Top panel:} Energy difference
  (absolute error) between MP2/6-311++G(d,p) reference energies and
  energies predicted by the NN trained on 100k reference
  structures. The error rarely exceeds 1 kcal
  mol$^{-1}$. \textit{Bottom panel:} The solid black curve corresponds
  to the reference energies, the dotted red curve corresponds to the
  energies predicted by the NN. It is able to describe transition
  geometries and geometries close to equilibrium structures equally
  well.}
\label{fig:ma}
\end{figure}

Illustrative results for reactions are presented for a NN using the
neighborhood representation for each atom for
malonaldehyde\cite{MM.nn:2018} and for PhysNet trained on S$_{\rm N}$2
reactions.\cite{unke2019physnet} A reactive MD simulation was run
using molecular mechanics with proton transfer\cite{MM.mmpt:2008} and
the energies were evaluated at the MP2/6-311++G(d,p) level of theory
and the NN trained on such energies. As is evident from figure
\ref{fig:ma}, the quality of the NN is better than 0.5 kcal/mol which
makes it competitive with an explicitly parametrized, reactive FF.\\

\begin{figure}[htbp]
\centering
\includegraphics[width=0.8\textwidth]{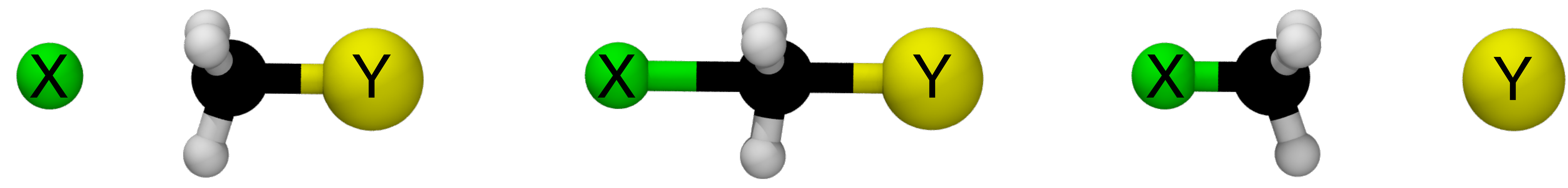}
\includegraphics[width=\textwidth]{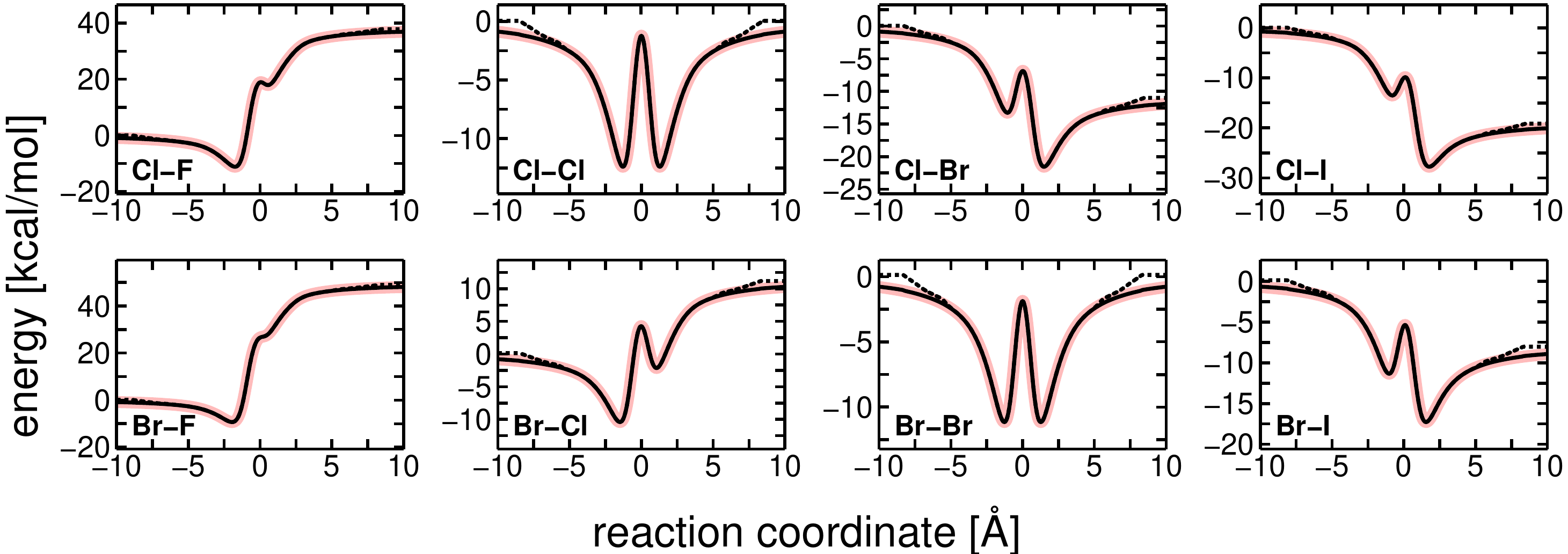}
\caption{Minimum energy paths (MEPs) for S$_{\rm N}$2 reactions
  X$^{-}$ + H$_3$C-Y $\rightarrow$ X-CH$_3$ + Y$^{-}$ along the
  reaction coordinate defined by the distance difference $r_{\rm
    CY}-r_{\rm CX}$, calculated using an ensemble of five NN models
  with (solid black line) and without (dotted black line) explicit
  long-range interactions. The the light red envelope is the MEP
  calculated using the reference method. Each panel shows a different
  combination X-Y, as indicated. The NN including explicit long-range
  interactions is virtually identical to the reference method for all
  values of the reaction coordinate (apart from small deviations in
  the asymptotics), whereas the model without long-range interactions
  shows qualitatively wrong asymptotic behaviour for a comparison of
  prediction errors between both models).}
\label{fig:sn2}
\end{figure}

\noindent
Using PhysNet the reactive PESs for S$_{\rm N}$2 reactions of the type
XCH$_3$--Y (with X and Y being F, Cl, Br, and I) were learned. As
reported in Figure \ref{fig:sn2} PhysNet without explicit long range
already performs well, in particular for the minimum and transition
state regions. However, including explicit long-range electrostatics
further enhances the performance of PhysNet which is also reflected in
reducing the MAE from 0.070 kcal/mol to 0.009 kcal/mol.

\section{Summary and Outlook}
Methods based on machine learning provide a versatile and
computationally robust and efficient framework to investigate the
dynamics of molecular systems. The present contribution briefly
discusses two avenues to generate high-accuracy representations based
on electronic structure calculations. One of them uses an
interpolation of gridded data based on reproducing kernels which is
also capable of describing the correct long range behaviour of
intermolecular interactions. The second approach (PhysNet) trains NNs
to reference electronic structure data in order to run reactive MD
simulations.\\

\noindent
One of the challenges for RKHS-based methods is to extend them to
higher-dimensional systems. Interesting work in this direction has
recently been discussed in the framework of Gaussian Process learning
which indicated that for accurate computation of differential cross
sections of a typical diatom-diatom system (O$_2$ + OH) only 300
energies are required to specify the PES.\cite{krems:2019} On the
other hand, work based on permutationally invariant polynomials was
able to parametrize a fully dimensional PES for
N-methyl-acetamide.\cite{bowman.nma:2019} Similar to a combination of
RKHS-based and empirical FFs together with MS-ARMD to combine accurate
representations of intermolecular interactions and the possibility to
follow chemical reactions it may be possible to combine NN-based
techniques with empirical FFs.  Another future challenge concerns
developing ML-learned representations for simulations in the condensed
phase.

\section*{Acknowledgments}
The authors acknowledge financial support from the Swiss National
Science Foundation (NCCR-MUST and Grant No. 200021-7117810), the
AFOSR, and the University of Basel.

\bibliography{references}

\end{document}